\documentclass[12pt]{iopart}
\usepackage{graphicx}
\usepackage{iopams}
\usepackage{amssymb}
\usepackage{multirow}

\newcommand{\ped}[1]{\ensuremath{_{\rm #1}}}

\begin{document}
\title{Strong-coupling Behaviour of Critical Temperature of Pb/Ag, Pb/Cu and Pb/Al Nanocomposites Explained by Proximity Eliashberg Theory}
\author{G.A.~Ummarino}
\ead{giovanni.ummarino@polito.it}
\address{Istituto di Ingegneria e Fisica dei Materiali,
Dipartimento di Scienza Applicata e Tecnologia, Politecnico di
Torino, Corso Duca degli Abruzzi 24, 10129 Torino, Italy; National Research Nuclear University MEPhI (Moscow Engineering Physics Institute),
Kashira Hwy 31, Moskva 115409, Russia}
\begin{abstract}

The experimental critical temperature of systems of nanoparticles superconducting ($Pb$) and normal ($Ag$, $Cu$ and $Al$) with a random distribution and sizes less than their respective coherence lengths, is governed by proximity effect as the experimental data shown. At first glance the behaviour of the variation of the critical temperature in function of the ratio of the volume fractions of the superconducting and the normal metal components seems to suggest a weak coupling behaviour for the superconductor. In reality, upon a more careful analysis using Eliashberg's theory for the proximity effect, the system instead shows a strong coupling nature.
The most interesting thing is that the theory has no free parameters and perfectly explains the behavior of the experimental data just with the assumption, in the cases of nanoparticles of $Ag$ and $Cu$, that the value of the density of states at the Fermi level of silver and copper is equal to value of lead.
\end{abstract}
%
{\textbf{Keywords:} Nanomaterials, Superconducting materials, Proximity effect, Eliashberg equations}
\maketitle
\section{INTRODUCTION}
\label{intro}
The superconductive proximity effect consist in the modification of the superconducting and normal properties of a superconductor in contact with a metal \cite{degennes,Wolf,revnostro,revsangita}. In the more general case the metal can be superconductive (with a lower critical temperature to respect to the first superconductor), normal or magnetic.
The superconducting coherence length $\xi_{S}=\frac{\hbar v_{FS}}{\pi \Delta}$ is the characteristic length scale of the proximity effect in the superconductor while in the metal is the coherence length of normal electrons $\xi_{N}=(\frac{\hbar v_{FN} l}{\pi k_{B} T})^{1/2}$, where $\Delta$ is the superconductive gap, $v_{FS,FN}$ is the Fermi speed of the superconductor and the normal metal, respectively, $l$ is the mean free path and $T$ the temperature. When the superconducting and normal layer thicknesses ($d_{S}$, $d_{N}$) are smaller than the respective coherence lengths $d_{S}<\xi_{S}$ and $d_{N}<\xi_{N}$ the system is the Cooper limit \cite{Alexp} ($S$ and $N$ indicate "superconductor" and "normal" respectively).
It is possible to demonstrate through experimental measurements and simple but accurate models in reproducing the aforementioned measurements that $T_{c}$ just depend on the thickness ratio $d_{S}/d_{N}$.
The original theory of the proximity effect in the framework before \cite{McMillan} of BCS theory and after \cite{Wolf,Carbi1,Carbi2,Carbi3,Carbi4,kresin,Ummaprox} of Eliashberg one usually presupposes a plain sample geometry i.e. the theory was born for describing the proximity effect between a slab of superconductor of thickness $d_{s}$ separated by a potential barrier from a slab of normal metal of
thickness $d_{N}$. A profound analogy exists between the proximity system and the two-gap model. If we assume that in the second band, as in the normal film, there is no intrinsic pairing (for example as it happen in the magnesium diboride \cite{MgB2a,MgB2b}) we have induced superconductivity i.e. an induced energy gap appears. The substantial difference in these two situations is
that the two-band model the bands are separated in momentum space and the second band acquires an order parameter due to phonon
exchange while in the proximity effect the systems are spatially separated, and superconductivity is induced by
the tunneling of Cooper pairs. In the first case the coupling is in the $k$-space while in the second is in the real space but the mathematical formalism is the same.
Furthermore also the effect of a static electric field on the critical temperature of a superconductor
can be explained in the framework of proximity Eliashberg theory \cite{Ummaprox1,Ummaprox2,Ummaprox3,Ummaprox4}.
Finally the role of Andreev reflection \cite{AR} is fundamentals in the undestanting
of microscopic mechanism at the origin of proximity effect \cite{Klap}. It happen that single electron states from normal metal are converted to Cooper pairs in the superconductor. The proximity effect can be seen as the result of interplay between long range order inside the normal metal and Andreev reflection at the normal metal-superconductor interface \cite{Pan}.
The link between Andreev reflection and proximity effect exist because the Andreev reflection of an electron or a hole is equivalent to the transfer of a Cooper pairs in or out of the superconductor i.e. to presence of Cooper pairs inside the normal metal.

Subsequently it was understood that in this theory could also be included the more general situation where just the ratio of the volume fraction of the superconducting and the normal metal components was known as happen in a two-component system consisting of a random distribution of superconducting and normal nanoparticles, with sizes less than their respective coherence lengths \cite{Alexp,Agexp,Cuexp}. In this case one may replace \cite{Alexp} the ratio of the thicknesses of the superconductor and normal metal layers $d_{S}/d_{N}$ in the de Gennes–Werthamer theory \cite{gennes1,gennes2} by the ratio of the volume fraction $P_{S}/P_{N}$.
This fact allows us to use the Eliashberg theory with proximity effect \cite{Ummaprox}, without free parameters, to explain the experimental data as well as has been done successfully for example with experimental data relative to $Pb/Ag$ heterostructure \cite{Nam} grown on Si(111) using molecular-beam epitaxy.
We will examine the cases where the superconducting nanoparticles are of lead while the normal ones are of silver, copper and aluminum.
According to the interpretation proposed by the authors of the measures \cite{Agexp} the lead behaves as a weak-coupling superconductor in the random lead-silver ($Pb-Ag$) nanocomposites . The same situation happens \cite{Cuexp} for the lead copper ($Pb-Cu$) nanocomposites. This interpretation is based on the use of analytical formulas obtained with numerous simplifications from the BCS theory and lead to non-physical values of some parameters. The problem is that the lead is a strong coupling superconductor and there is no reason why it should behave differently.
In the last case \cite{Alexp}, which will be analyzed, lead-aluminum ($Pb-Al$) nanocomposites, the strong coupling behavior will be immediately evident.
In principle one should also consider in which way the electron-phonon interaction interplays with the quantum-size effects \cite{Cro1,Cro2}. Here we are not dealing with individual grains but rather nanocrystalline films where, probably, these effects, which on the critical temperature can be of the opposite sign, on average, cancel each other out.
The paper is organized as follow. In Sec.~\ref{sec:model} we show the model we use for the computation of the superconductive critical temperature of these proximity system, i.e. the one band s-wave Eliashberg equations with proximity effect. In Sec.~\ref{sec:results} we discuss my results in comparison with experimental data. Finally, conclusions are given in Sec.~\ref{sec:conclusions}.
\section{MODEL: PROXIMITY ELIASHBERG EQUATIONS}
\label{sec:model}
By solving the one band s-wave Eliashberg equations \cite{carbibastardo,ummarinorev}, generalized to case where is present the proximity effect, it is possible to calculate the critical temperature of this superconducting-normal system. Four coupled equations: two for the gaps $\Delta_{S(N)}(i\omega_{n})$ and two for the renormalization functions $Z_{S(N)}(i\omega_{n})$ have to be solved. If the Migdal theorem is valid \cite{Migdal}, the Eliashberg equations with proximity effect on the imaginary-axis \cite{Carbi1,Carbi2,Carbi3,Carbi4,kresin,Ummaprox,Ummaprox1}($\omega_{n}$ denotes the Matsubara frequencies), for the gaps $\Delta_{S,N}(i\omega_{n})$, read:
\begin{eqnarray}
&&Z_{N}(i\omega_{n})\Delta_{N}(i\omega_{n})=\pi
T\sum_{m}\big[\Lambda_{N}(i\omega_{n},i\omega_{m})-\mu^{*}_{N}(\omega_{c})\big]\times\nonumber\\
&&\times\Theta(\omega_{c}-|\omega_{m}|)N^{\Delta}_{N}(i\omega_{m})
+\Gamma\ped{N} N^{\Delta}_{S}(i\omega_{n})\phantom{aaaaaa}
 \label{eq:EE2}
\end{eqnarray}
\begin{eqnarray}
&&Z_{S}(i\omega_{n})\Delta_{S}(i\omega_{n})=\pi
T\sum_{m}\big[\Lambda_{S}(i\omega_{n},i\omega_{m})-\mu^{*}_{S}(\omega_{c})\big]\times\nonumber\\
&&\times\Theta(\omega_{c}-|\omega_{m}|)N^{\Delta}_{S}(i\omega_{m})
+\Gamma\ped{S}N^{\Delta}_{N}(i\omega_{n})\phantom{aaaaaa}
 \label{eq:EE4}
\end{eqnarray}
while ones for the renormalization functions $Z_{S,N}(i\omega_{n})$ are
\begin{eqnarray}
&&\omega_{n}Z_{N}(i\omega_{n})=\omega_{n}+ \pi T\sum_{m}\Lambda_{N}(i\omega_{n},i\omega_{m})N^{Z}_{N}(i\omega_{m})+\nonumber\\
&&+\Gamma\ped{N} N^{Z}_{S}(i\omega_{n})
\label{eq:EE1}
\end{eqnarray}
\begin{eqnarray}
&&\omega_{n}Z_{S}(i\omega_{n})=\omega_{n}+ \pi T\sum_{m}\Lambda_{S}(i\omega_{n},i\omega_{m})N^{Z}_{S}(i\omega_{m})+\nonumber\\
&&\Gamma\ped{S} N^{Z}_{N}(i\omega_{n})
\label{eq:EE3}
\end{eqnarray}
where $\Theta$ is the Heaviside function, $\omega_{c}$ is a cutoff energy and $\mu^{*}_{S(N)}(\omega_{c})$ are the Coulomb pseudopotentials in the superconductive and normal layer respectively.
The coupling terms between normal and superconductive layers are
\begin{equation}
\Gamma_{S(N)}=\pi|t|^{2}Ad_{N(S)}N_{N(S)}(0)
\label{eq:EE6}
\end{equation}
 where $|t|^{2}$ is the transmission matrix, $A$ is the junction cross-sectional area, $d_{S(N)}$ are the superconductive and normal layer thicknesses respectively, $N_{S(N)}(0)$ are the densities of states atthe Fermi level for the superconductive and normal material and the rate of the coupling terms is $\frac{\Gamma_{S}}{\Gamma_{N}}=\frac{d_{N}N_{N}(0)}{d_{S}N_{S}(0)}$.
Finally the terms relative to quasiparticles and Cooper pair density of states are:
\begin{equation}
N^{Z}_{S(N)}(i\omega_{m})=\omega_{m}/{\sqrt{\omega^{2}_{m}+\Delta^{2}_{S(N)}(i\omega_{m})}}
\end{equation}
\begin{equation}
N^{\Delta}_{S(N)}(i\omega_{m})=\Delta_{S(N)}(i\omega_{m})/
{\sqrt{\omega^{2}_{m}+\Delta^{2}_{S(N)}(i\omega_{m})}}
\end{equation}
The phononic glue is inside the following term:
\begin{equation}
\Lambda_{S(N)}(i\omega_{n},i\omega_{m})=2
\int_{0}^{+\infty}d\Omega \Omega
\alpha^{2}_{S(N)}F(\Omega)/[(\omega_{n}-\omega_{m})^{2}+\Omega^{2}]
\end{equation}
where $\alpha^{2}_{S(N)}F(\Omega)$ are the electron-phonon spectral functions and the electron-phonon coupling constants are defined as
\begin{equation}
\lambda_{S(N)}=2\int_{0}^{+\infty}d\Omega\frac{\alpha^{2}_{S(N)}F(\Omega)}{\Omega}\nonumber\\
\end{equation}
In this set of coupled equations appears a lot of input parameters but, luckily, are all known because these materials are phononic. In particular these input parameters are: two electron-phonon spectral functions $\alpha^{2}_{S(N)}F(\Omega)$, two Coulomb pseudopotentials $\mu^{*}_{S(N)}(\omega_{c})$, two values of the normal density of states at the Fermi level $N_{S(N)}(0)$ and the thickness of the superconductive layer $d_{S}$ and normal one $d_{N}$ (these last are experimental inputs). In principle also the product between the junction cross-sectional area $A$ and transmission matrix $|t|^{2}$ are present but we have verified, as it should be, that the final result does not depend on the value of $A$ and therefore not even on the value of the product $A|t|^{2}$.

All inputs parameters are known and will be specified below.
The letter $S$ is for $Pb$ and the letter $N$ is for $Ag$, $Cu$ and $Al$.
The electron-phonon spectral functions $\alpha^{2}_{S(N)}F(\Omega)$ of lead ($\lambda_{Pb}=1.55$) \cite{carbibastardo}, aluminum ($\lambda_{Al}=0.43$) \cite{a2fAl}, silver ($\lambda_{Ag}=0.16$) \cite{a2fAgCu} and copper ($\lambda_{Cu}=0.14$) \cite{a2fAgCu} are present in literature (they are shown in the insert of the figures) as the value of the normal density of states at the Fermi level \cite{Ummaprox1,ButlerN0AgCuAl} $N_{Pb}(0)=0.25866$ $eV^{-1}$ for unit cell, $N_{Al}(0)=0.20000$ $eV^{-1}$ for unit cell, $N_{Ag}(0)=0.13000$ $eV^{-1}$ for unit cell and $N_{Cu}(0)=0.13000$ $eV^{-1}$ for unit cell.
The value of the Coulomb pseudopotential is fixed for obtaining $T_{c}=7.20$ $K$ in a system without proximity effect for lead and $T_{c}=1.18$ $K$ for aluminum: we find $\mu^{*}_{Pb}(\omega_{c})=0.14023$ and $\mu^{*}_{Al}(\omega_{c})=0.14448$ by using a cutoff energy $\omega_{c} = 125$ meV and a maximum energy $\omega_{max} = 130$ meV. The Coulomb pseudopotential for copper and silver is the same \cite{a2fAgCu} $\mu^{*}_{Ag,Cu}(\omega_{c})=0.11000$. The values of $d_{S}$ and $d_{N}$ are experimental data. By specifying these inputs parameters this calculation has no free parameters.

\section{RESULTS AND DISCUSSION}
\label{sec:results}
The Eliashberg equation are solved numerically in recursive way. This is a standard method that work very well because quickly the solution is reached \cite{carbibastardo}.
From the solution of  Eliashberg equations it is possible to determine the critical temperature as a function of the ratio of the volume fraction of the superconducting and the normal metal components $P_{S}/P_{N}$. But the comparison with experimental data is not so good as seen in figure 1 and figure 2 (dark blue solid line). So it is necessary to do some new hypothesis to solve this problem, precisely one.
We assume that, to explain the experimental data, it is necessary that the density of states at the Fermi level of the
normal metal nanoparticles has to be substitute, in the equations, by the value of the superconductor density of states
but only when the size of normal metal nanoparticles is minor to that of the superconductor coherence length. The characteristic length of the proximity effect is the coherence length that is the typical size of the Cooper pairs and the proximity effect is connected with the Cooper pairs inside the normal metal. If the dimensions of the nanoparticles are only smaller than the coherence length of the superconductor (i.e. the dimensions of the Cooper pairs) it is possible that the electronic properties of the nanoparticles are replaced by those of the superconductor.
The coherence length of lead \cite{coherencePb} is $96$ $nm$ so the silver and copper nanoparticle size is always less than this distance.
In this way we change just the value of $N_{Ag,Cu}(0)$: now $N_{Ag,Cu}(0)=N_{Pb}(0)$ but all other input parameters remain the same and we solve the Eliashberg equations. The result is the solid red line in figures 1 and 2 that shows a very good agreement with experimental data. The author found the same behaviour in the superconductor/normal metal heterostructure $(Pb/Ag)$, epitaxially grown \cite{Nam}. Also in this case we had made the same hypothesis and had perfectly reproduced the experimental data \cite{Ummaprox}. One could say that within the superconducting coherence length the superconductor "wins" over the metal and this is observable because lead has a large coherence length. Since the same phenomenon of a faster, than expected, decrease of the critical temperature of these systems have occurred both in nanocomposites and in high quality superconductor/normal metal heterostructure, the reason it will be of a general nature and will probably not concern the quality or the particular characteristics of the samples. So this assumption ($N_{Ag,Cu}(0)=N_{Pb}(0)$) that allows to explain very well the experimental data could lead to investigating the nature of these systems using first-principles calculus.
If the Eliashberg theory is assumed to be valid, and we see no reason to doubt it, the only way to reproduce the experimental data is by this assumption. The only other input parameters which, in theory, could be changed are the Coulomb pseudopotential and the electron-phonon coupling constants but in this case we should admit that these input parameters are a function of $P_{S}/P_{N}$ because for high values of $P_{S}/P_{N}$ we have to regain the lead bulk critical temperature. Furthermore even by varying the Coulomb pseudopotential it is not possible to reproduce the experimental data in any way as we have been able to verify.

In the last case (aluminum) the standard theory without free parameters reproduces the experimental data very well as it is shown in figure 3. In the latter case, the aluminum which is a superconductor, also if in the temperature range studied is a normal metal, follows perfectly the standard theory. For completeness, the calculation is also shown in the figure with the replacement of the density of the normal states with the superconducting one as in the previous cases. Here it is very clear that this method doesn't work. The reason is simple: in this case the standard theory works because the Al layer size is greater and therefore the starting assumption is no longer valid.
\section{CONCLUSIONS}
\label{sec:conclusions}
The experimental critical temperature of systems of nanoparticles superconducting ($Pb$) and normal ($Ag$, $Cu$) with a random distribution can be reproduced very well in the framework of Eliashberg theory by assuming that, the density of the states at the Fermi level of the superconductor is replaced by the value present in the normal material in the proximity systems $Pb-Ag$ and $Pb-Cu$.
In the last case ($Pb-Al$) the standard theory explain perfectly the experimental data without an additional hypothesis and, of course, free parameters. We emphasize that to justify this assumption it would be necessary to resort to calculations from first principles which are not present in the literature.
In general it is possible to state that all the experimental data are accurately described by Eliashberg's theory of proximity effect with no free parameters and without no "strange weak coupling behaviour"
as had been hypothesized by the authors of the measurements on the $Pb-Ag$ proximity system \cite{Agexp}.

G.A. Ummarino thanks for the support received from the MEPhI Academic Excellence Project(Contract No. 02.a03.21.0005).\\


\newpage
\begin{figure}
\begin{center}
\includegraphics[keepaspectratio, width=0.9\columnwidth]{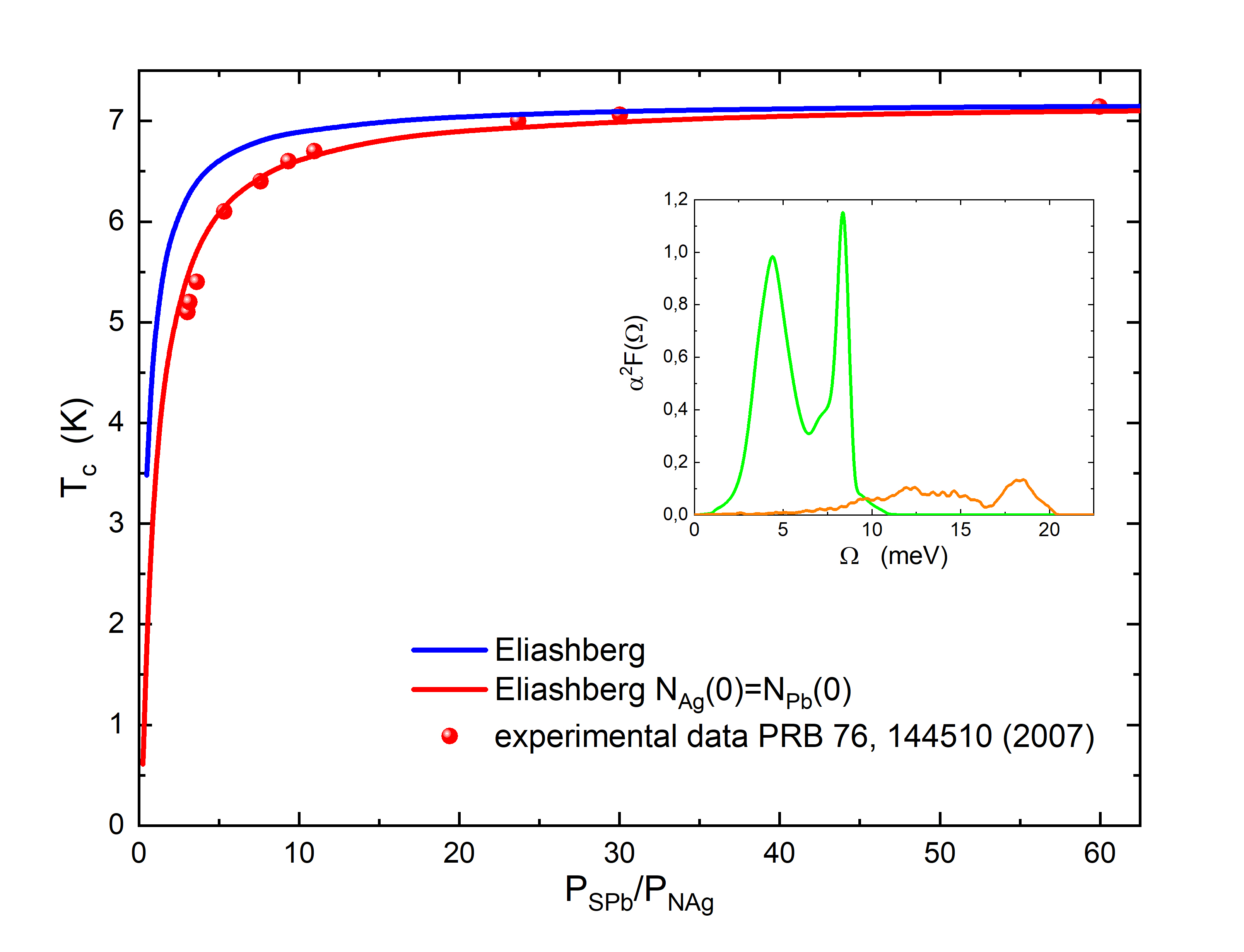}
\vspace{-5mm} \caption{(Color online)
 Case $Pb-Ag$. Theoretical critical temperature calculated by solving the Eliashberg equations with $N_{Ag}(0)\neq N_{Pb}(0)$ (dark blue solid line) and with $N_{Ag}(0)=N_{Pb}(0)$ (red solid line) in function of the rate $P_{S}/P_{N}$ is shown. The experimental data (full red circles) are from ref \cite{Agexp}. In the insert it is shown the electron-phonon spectral functions of lead (green solid line) and silver (orange solid line).
 }\label{Figure1}
\end{center}
\end{figure}
\newpage
\begin{figure}
\begin{center}
\includegraphics[keepaspectratio, width=0.9\columnwidth]{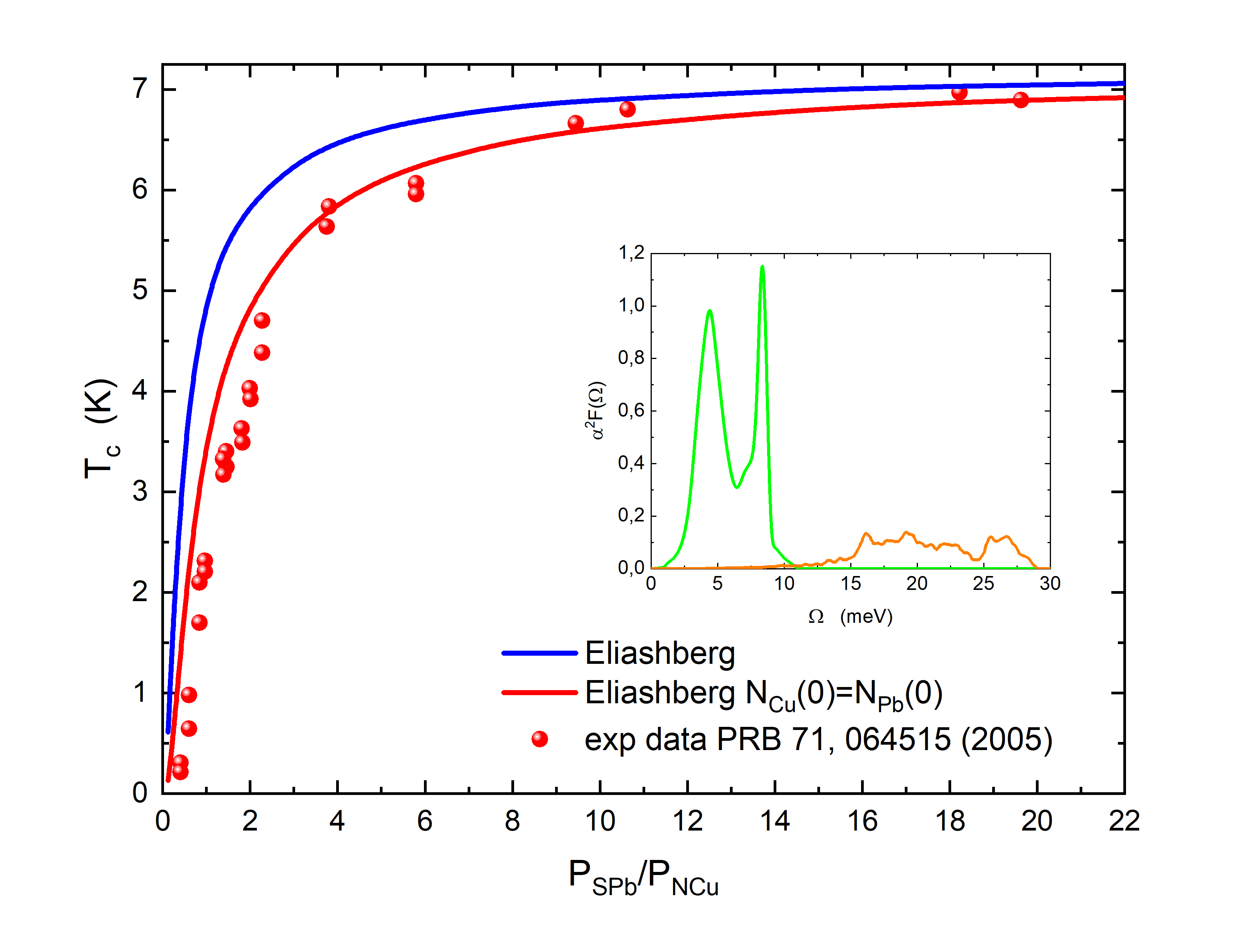}
\vspace{-5mm} \caption{(Color online)
 Case $Pb-Cu$. Theoretical critical temperature calculated by solving the Eliashberg equations with $N_{Cu}(0)\neq N_{Pb}(0)$ (dark blue solid line) and with $N_{Cu}(0)=N_{Pb}(0)$ (red solid line) in function of the rate $P_{S}/P_{N}$ is shown. The experimental data (full red circles) are from ref \cite{Cuexp}. In the inset it is shown the electron-phonon spectral functions of lead (green solid line) and copper (orange solid line).
 }\label{Figure2}
\end{center}
\end{figure}
\newpage
\begin{figure}
\begin{center}
\includegraphics[keepaspectratio, width=0.9\columnwidth]{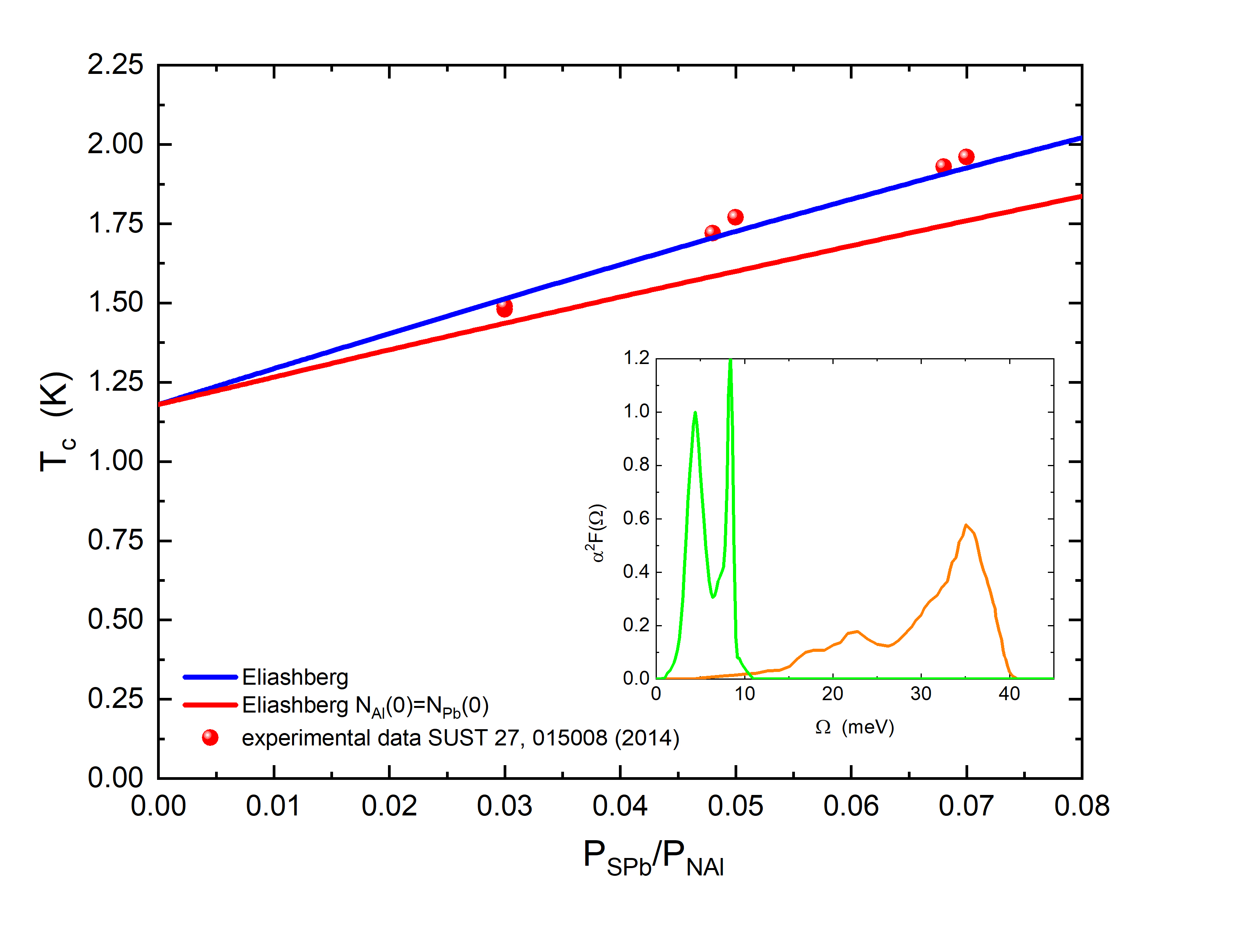}
\vspace{-5mm} \caption{(Color online)
 Case $Pb-Al$. Theoretical critical temperature calculated by solving the Eliashberg equations with $N_{Al}(0)\neq N_{Pb}(0)$ (dark blue solid line) and with $N_{Al}(0)=N_{Pb}(0)$ (red solid line) in function of the rate $P_{S}/P_{N}$ is shown. The experimental data (full red circles) are from ref \cite{Alexp}. In the inset it is shown the electron-phonon spectral functions of lead (green solid line) and aluminum (orange solid line).
 }\label{Figure3}
\end{center}
\end{figure}
\end{document}